\documentclass{article}
\usepackage{amsmath,geometry,physics,amssymb,mathtools,array,tabularx,amsfonts,cite,tcolorbox,subcaption,tocloft,enumitem,textcomp,tikz,jheppub}

\newcommand{\cl}{\text{cl}}
\newcommand{\UV}{\text{UV}}

\newcommand{\zz}{\mathtt{z}}

\definecolor{DCviolet}{RGB}{140, 43, 226}
\newcommand{\DC}[1]{{\color{DCviolet}{DC: #1}}}

\begin{document}

\title{Holographic Spread Complexity from Branes and Strings}

\author[a]{Dimitrios Chatzis,}\author[a]{ Madison Hammond,} \author[a]{Carlos Nunez,}\author[b]{Alfonso V. Ramallo}  \author[a]{ and ~~~~~~~~~~~Ricardo T. Santamaria}
\affiliation[a]{Centre for Quantum Fields and Gravity, Department of Physics, Swansea University, Swansea SA2 8PP, United Kingdom}
\affiliation[b]{Departamento de F\'{\i}sica de Part\'{\i}culas, Universidade de Santiago de Compostela and Instituto Galego de F\'{\i}sica de Altas Enerx\'{\i}ıas (IGFAE). E-15782 Santiago de Compostela, Spain}
%
%
\abstract{ 
We study Krylov spread complexity in holographic theories using genuine string-theory probes. Building on the proposal that the growth rate of spread complexity is measured by a proper momentum in the bulk, we embed the falling-particle picture in top-down examples. We first analyse a D0 brane in the type IIA AdS$_4\times {\mathbb{CP}}^3$ background dual to ABJM theory, identifying it with a dressed monopole operator in the boundary CFT. For purely radial motion the proper-momentum prescription reproduces the expected quadratic growth of the complexity. When the probe carries momentum along an isometric direction, the naive prescription gives an apparent conflict with the short-time behaviour required of Krylov complexity. We propose that the correct fixed-charge description is obtained by Legendre transforming to the Routhian.
We support the D0-brane interpretation through the regulated monopole two-point function, whose survival amplitude determines the Krylov moments, and we show that radial fluctuations give controlled corrections to the effective energy governing the complexity growth. We then extend the analysis to a rotating non-BPS D3 brane in AdS$_5\times S^5$, where angular momentum produces a centrifugal barrier and a sharp condition for radial in-fall. In the falling regime the Routhian prescription again gives the correct short-time behaviour. Finally, we consider a wound fundamental string in AdS$_5\times S^5$, which reduces to an effective massive falling particle. This clarifies the distinction between Noether charges, which require a fixed-charge Routhian treatment, and winding data, which enter through the effective mass. Our results provide a string-theoretic realisation of holographic spread complexity for point-like and extended excitations, making manifest their dependence on field theory parameters.
}
\emailAdd{dchatzis@proton.me}
\emailAdd{m.hammond.2412736@swansea.ac.uk}
\emailAdd{c.nunez@swansea.ac.uk}
\emailAdd{alfonso.ramallo@usc.es}
\emailAdd{ricardo.sta2718@gmail.com}
\maketitle
\flushbottom

\section{Introduction and General Idea of this paper}

The notion of quantum complexity has become a useful language for quantifying how a state, or an operator, explores the Hilbert space available to it.  In holography, this question is especially suggestive: the time evolution of apparently simple boundary data can be mapped to geometric motion in the bulk, and one may ask which bulk observable measures the growth of the corresponding boundary complexity.  In this paper we study this question for a class of top-down string-theory probes.  Our aim is not to introduce a new abstract definition of complexity, but rather to sharpen the holographic interpretation of Krylov spread complexity for genuine stringy excitations of well-defined AdS/CFT dual pairs.

Let us first briefly recall the Krylov construction for states.  Given a Hamiltonian $H$ and an initial state $|\psi_0\rangle$, one builds an orthonormal Krylov basis $\{|K_n\rangle\}$ by applying the Lanczos algorithm to the sequence $|\psi_0\rangle,H|\psi_0\rangle,H^2|\psi_0\rangle,\ldots$.  Equivalently, the dynamics is encoded in the survival amplitude
\begin{equation}
        A(t)=\langle \psi_0|e^{-iHt}|\psi_0\rangle
        =\int dE\,\rho(E)e^{-iEt},
\end{equation}
whose Taylor coefficients of the time expansion, determine the moments of the spectral measure.  From these moments one obtains the Lanczos coefficients, the Krylov wavefunctions $\phi_n(t)$ defined by
\begin{equation}
        e^{-iHt}|\psi_0\rangle=\sum_{n\geq 0}\phi_n(t)|K_n\rangle,
\end{equation}
and finally the spread complexity
\begin{equation}
        {\cal C}(t)=\sum_{n\geq 0} n\,|\phi_n(t)|^2 .
\end{equation}
This quantity measures the average position of the evolved state along the Krylov chain and is therefore a direct measure of how efficiently the time evolution spreads the initial state over the basis dynamically generated by $H$.  We shall use the standard spread-complexity technology developed in \cite{Balasubramanian:2022tpr,
Caputa:2023vyr,
Balasubramanian:2025xkj,
Caputa:2025ozd}.  In particular, we will repeatedly use the fact that, for a unitary time evolution generated by a Hermitian Hamiltonian, ${\cal C}(t)$ is an even function of time and hence its short-time expansion starts as ${\cal C}(t)\sim t^2$, or equivalently $\dot{\cal C}(t)\sim t$ \cite{Muck:2026top}.

The bridge to holography that motivates this work is the proper-momentum proposal of \cite{Caputa:2024sux}.  For states created by local operators in a CFT, the rate of spread complexity can be matched to the proper radial momentum of a massive particle falling in the dual AdS spacetime.  Schematically, after choosing the appropriate proper coordinate $y$ along the bulk trajectory, the proposal takes the form
\begin{equation}
        \dot{\cal C}(t)\sim |P_y(t)| .
\end{equation}
This idea has been developed and extended in several directions, including refinements of the proper-momentum prescription \cite{Fan:2024iop,He:2024pox}, top-down examples involving quiver geometries and internal directions \cite{Fatemiabhari:2025cyy, Fatemiabhari:2025poq,Fatemiabhari:2025usn,Fatemiabhari:2026goj, Zoakos:2026obl, Roychowdhury:2026igc, Roychowdhury:2026sgg, Alfinito:2026vah, Alfinito:2026yex} and charged or composite probes in ${\cal N}=4$ SYM and related systems \cite{Nastase:2026lhz,ChHNRS}.  These works suggest that Krylov complexity is sensitive not only to radial infall, but also to additional structure of the dual operator: R-charges, motion in internal spaces, quiver directions, compositeness and extendedness.  This is the perspective we adopt below.

There is, however, a conceptual point that becomes unavoidable once the probe carries a conserved Noether charge.  If the bulk excitation moves along an isometric direction, the naive proper momentum can contain a constant contribution already at $t=0$.  Interpreted directly as $\dot{\cal C}$, this would contradict the evenness of the Krylov complexity \cite{Muck:2026top}.  The central proposal of the present paper is that the correct holographic prescription in such a sector is a fixed-charge prescription: one should Legendre transform with respect to the cyclic coordinate and work with the corresponding Routhian.  The proper momentum is then computed from the fixed-charge dynamics.  This restores the expected short-time behaviour whilst preserving the physical effect of the conserved charge.  In this sense, the Routhian is not just a calculational convenience; it is the natural bulk implementation of working in a definite charge sector of the boundary theory.  This viewpoint is also compatible with recent developments on symmetry-resolved and multi-seed versions of Krylov complexity \cite{Das:2024tnw,Craps:2024suj,Caputa:2025mii,Caputa:2025ozd}.

The purpose of this paper is to test this idea in examples where the bulk excitation is not an abstract massive particle, but a genuine object in the string spectrum.  In Section~\ref{SectionABJMm} we begin with the ABJM duality \cite{Aharony:2008ug}.  We consider the type IIA AdS$_4\times \mathbb{CP}^3$ background, reviewed for our conventions in \cite{Bea:2013jxa}, and study a falling D0 brane.  This is the most direct top-down realisation of the particle considered in \cite{Caputa:2024sux}: the D0 brane is a pointlike bulk object, but it is also a genuine type IIA excitation and is naturally associated with monopole operators in the ABJM CFT.  The relevant dressed monopole operators and their role in the ABJM spectrum were discussed in \cite{Benna:2009xd,Berenstein:2009sa}.  Thus the gravitational trajectory is tied to a concrete boundary state, rather than to a phenomenological point particle. Sections \ref{sectionD3branesinAdS} and \ref{seccion-4}, extend this treatment to other genuine excitations of AdS$_5\times S^5$.

In more detail, the contents of this work are distributed as follows.
\begin{itemize}
\item{Section~\ref{SectionABJMm} contains several related results.  Firstly, for a D0 brane falling with no spatial momentum, the proper-momentum prescription reproduces the standard linear growth of $\dot{\cal C}$ and therefore ${\cal C}\sim t^2$.  Secondly, when the D0 brane carries momentum along a boundary spatial isometry, the naive prescription produces the fixed-charge puzzle described above.  We resolve it by Legendre transforming with respect to the isometric coordinate and computing the proper momentum from the Routhian.  The resulting rate of complexity has the correct short-time behaviour, reduces smoothly to the uncharged result, and retains a non-trivial dependence on the conserved momentum.  Thirdly, using the regulated two-point function of the monopole operator, we identify the survival amplitude $A(t)$
from which the moments $\mu_k$ can be read explicitly.  This gives a direct field-theory route from CFT data to Lanczos coefficients, Krylov wavefunctions and spread complexity.  Finally, we include radial fluctuations of the D0 probe.  These fluctuations represent internal stringy dynamics beyond the rigid-particle approximation and give controlled perturbative corrections to the rate of complexity, which at leading order can be interpreted as a shift of the effective energy of the falling probe.}
\item{
In Section~\ref{sectionD3branesinAdS} we move from pointlike probes to extended objects by considering a D3 brane in $\mathrm{AdS}_5\times S^5$.  The brane extends along the Minkowski directions, falls in the AdS radial direction and rotates on the internal $S^5$.  This angular velocity makes the configuration non-BPS and introduces a conserved angular momentum.  Related brane probes and giant-graviton-type configurations have been studied in several contexts, see for example \cite{Das_2000,Kim_2001, Camino:2001ti}, but here the emphasis is on their role as extended operators whose spread complexity is affected by internal motion.  We derive the first-order dynamics, obtain a clean falling criterion, and rewrite the radial motion as that of a particle of zero energy in an effective potential.  This makes the physics transparent: for certain parameter ranges the D3 brane falls, while outside them a centrifugal barrier prevents radial infall (and the associated complexity growth).

The D3 example provides a second and more stringent test of the fixed-charge prescription.  Working at fixed angular momentum, we construct the Routhian and define the associated proper coordinate and proper momentum.  In dimensionless variables this leads to a particularly simple expression for $P_y$ and hence for $\dot{\cal C}$.  The result again has the expected short-time behaviour, while at late times it displays a different growth from the uncharged particle case.  The lesson is that the complexity of an extended non-BPS operator is sensitive to more data than its total energy: it also knows about the charge sector, the existence of a centrifugal barrier, and the distinction between falling and non-falling trajectories.}

\item{In Section~\ref{seccion-4} we study a complementary string example in the same $\mathrm{AdS}_5\times S^5$ geometry.  Starting from the Polyakov action, we identify a consistent truncation describing a fundamental string that falls in the radial direction while winding internal angular directions.  On this solution the Nambu--Goto action reduces to the action of an effective massive particle in AdS, with a mass controlled by the fundamental-string tension, the AdS radius, the length of the $\sigma$-direction and the winding number.  The proper-momentum prescription then reproduces the familiar quadratic growth ${\cal C}\sim t^2$, but now for a genuine extended string excitation.  This example is conceptually useful because the winding number is a conserved topological datum rather than a Noether charge requiring non-zero initial velocity.  Therefore no Routhian treatment is needed in this sector.  More general string solutions carrying Noether charges are expected to combine the two mechanisms: effective-mass renormalisation due to winding and fixed-charge Routhian dynamics due to motion along isometries.}
\end{itemize}
The common message of the three examples is that holographic spread complexity is a sensitive diagnostic of the microscopic nature of the excitation.  The point-particle result of \cite{Caputa:2024sux} is recovered in the appropriate limit, but genuine stringy probes bring new structure: monopole quantum numbers in ABJM, radial fluctuations, angular momenta, centrifugal barriers, extended brane dynamics and winding sectors.  These effects suggest a broader organisation of holographic Krylov complexity in which the collective radial motion gives one contribution, while charges, internal excitations and extended degrees of freedom give additional sectors.  The conclusions collect these lessons and indicate how the fixed-charge prescription may be used in more general top-down constructions.
\section{D0 branes in ABJM}\label{SectionABJMm}
In this section we put on firm 'string-theoretic-basis' some of the contents in \cite{Caputa:2024sux, Fan:2024iop, He:2024pox}. We probe AdS-space (in our case the ABJM background \cite{Aharony:2008ug}) with a point-like excitation, represented by a D0 brane. We identify this probe with a dressed monopole operator in the three dimensional ABJM SCFT.

Let us consider the type IIA solution found by dimensionally reducing M-theory on $\mathrm{AdS}_4\times \mathbb{S}^7/\mathbb{Z}_k$ \cite{Aharony:2008ug}. The solution is comprised of a metric (in string frame and using units with $\alpha'=1$),

\begin{equation}\label{AdS4xCP3}
    \begin{split}
        &\dd s^2_{10}=L^2_{\text{ABJM}}(\dd s^2_{\text{AdS}_4}+4 \dd s^2_{\mathbb{CP}^3}),\\
        &\dd s^2_{\text{AdS}_4}=r^2 (-\dd t^2+\dd x_1^2+\dd x_2^2)+\frac{\dd r^2}{r^2},\\
        &\dd s^2_{\mathbb{CP}^3}=g_{ij}(\zz_l,\bar{\zz}_l)\dd \zz^i \dd \bar{\zz}^j .
    \end{split}
\end{equation}
We do not quote the specific form of the metric on $\mathbb{CP}^3$, as it is not used in this work. For details see Section 2 of \cite{Bea:2013jxa}.
There is a constant dilaton, given by
\begin{equation}
    e^\Phi= \frac{2}{k}L_{\text{ABJM}}=2\sqrt{\pi} \left( \frac{2N}{k^5}\right)^{1/4},~~\text{with}~~  L_{\text{ABJM}}^4=2\pi^2 \frac{N}{k}.
\end{equation}
The solution is complemented by Ramond fields $F_2$ and $F_4$, that we do not quote--see \cite{Bea:2013jxa} for details. We use the above metric to study geodesics of a falling D0 brane.
%
We consider the embedding parametrised in terms of the $t$-coordinate,
\begin{equation}
   \quad x_i=x_i(t),\quad r=r(t),\quad \zz ^i=\zz ^i (t).
\end{equation}
This embedding is associated with the induced metric,
\begin{equation}
    \dd s^2_{\text{ind}}=L^2_{\text{ABJM}} \Big[ r^2(\dot{x}_1^2+\dot{x}_2^2-1)+\frac{\dot{r}^2}{r^2}+ 4g_{ij}\dot{\zz}^i\dot{\bar{\zz}}^j \Big]\dd t^2,
\end{equation}
and the  action for the probe D0 brane is 
\begin{equation}
    \begin{split}
        S_{\text{DBIWZ}}&=\int \dd t ~e^{-\Phi}\sqrt{-\text{det}g_{\text{ind}}}+\int P[C_1]\\
        &=\frac{k}{2}\int \dd t \left( \sqrt{ r^2(1-\dot{x}_1^2 - \dot{x}_2^2) - \frac{\dot{r}^2}{r^2}-4 g_{ij}\dot{\zz}^i \dot{\bar{\zz}}^j}- \tilde{C}_{1,i}\dot{\zz}^i \right).
    \end{split}
\end{equation}
The conjugate momenta for the embedding coordinates are given by
\begin{equation}
    \begin{split}
       & {\cal P}_{x_{1,2}}=-\frac{k}{2}\frac{r^2\dot{x}_{1,2}}{\sqrt{{\cal D}}},\quad {\cal P}_{r}=-\frac{k}{2}\frac{\dot{r}}{r^2\sqrt{{\cal D}}},\quad {\cal P}_{\zz^i}=-\frac{k}{2}\left( \frac{4g_{ij}\dot{\bar{\zz}}^j}{\sqrt{{\cal D}}}+\tilde{C}_{1,i}\right),\\
       &\text{where}\quad {\cal D}=r^2(1-\dot{x}_1^2-\dot{x}_2^2)-\frac{\dot{r}^2}{r^2}-4g_{ij}\dot{\zz}^i\dot{\bar{\zz}}^j.
    \end{split}
\end{equation}
We discuss the case of a D0 probe that falls keeping constant $x_2$ (without loss of generality). The $\zz^i$-coordinates are also set to constants (embeddings without motion in $\mathbb{CP}^3$).
\subsection{D0 with~ \texorpdfstring{$x_1(t)\,~ \text{and} ~ \, r(t)$}{x(t) and r(t)}}
As anticipated, we focus on a simple embedding, describing a D0 with a profile in the radial and one spatial ${\rm{AdS}}$ directions,
\begin{equation}
    \text{D}0:\quad r=r (t) \quad \&\quad x_1=x_1(t).
\end{equation}
We keep the internal coordinates $\zz^i$ to be suitably determined constants by solving their equations of motion. The induced metric, Lagrangian density, conjugate momenta and Hamiltonian for the D0-probe read
\begin{eqnarray}
& & 
\dd s^2_{\text{ind}}=L_{\text{ABJM}}^2\left[ r^2(-1+\dot{x}_1^2) + \frac{\dot{r}^2}{r^2}\right]\dd t^2,\quad 
    {\cal L}=\frac{k}{2}\sqrt{r^2(1-\dot{x}_1^2)-\frac{\dot{r}^2}{r^2}},\label{metric_and_lagrangian_D0}\\
    & & {\cal P}_{x_1} =- \frac{k}{2}\frac{r^3\dot{x}_1}{\sqrt{r^4(1-\dot{x}_1^2)-\dot{r}^2}},\quad {\cal P}_r = -\frac{k}{2}\frac{\dot{r}}{r\sqrt{r^4(1-\dot{x}_1^2)-\dot{r}^2}},\nonumber\\
       & &{\cal H} = \frac{k}{2}\frac{r^3}{\sqrt{r^4 (1-\dot{x}_1^2) - \dot{r}^2}}.\nonumber
\end{eqnarray}
%
%

The conjugate momentum associated with the coordinate $x_1$ and the Hamiltonian are both conserved. Note that we defined the energy of the D0 brane to be ${\cal H}=-H_{\text{canonical}}$. In terms of these quantities, the velocity in the $r$-direction reads
\begin{equation}
    \dot{r}=- \frac{r^2}{2{\cal H}}\sqrt{4({\cal H}^2-{\cal P}_{x_1}^2)-k^2 r^2}.\label{eqdotr}
\end{equation}
We have chosen the minus sign to indicate that the particle falls (negative $\dot{r}$). Imposing an initial condition  for $t=0$,  $r(0)=r_{\text{UV}}$ and $\dot{r}(0)=0$ sets
\begin{equation}
    r _{\text{UV}} = \frac{2}{k}\sqrt{{\cal H}^2 - {\cal P}_{x_1}^2}.
\end{equation}
We solve for $r(t)$ and $x_1(t)$, 
\begin{equation}\label{sols_x1_r}
         x_1 (t) = -\frac{{\cal P}_{x_1}}{{\cal H}}t + x_{1,0},~~~~
         r(t) =  \frac{2{\cal H}\sqrt{{\cal H}^2 - {\cal P}_{x_1}^2}}{\sqrt{k^2 {\cal H}^2 +4 ({\cal H}^2 - {\cal P}_{x_1}^2)^2 t ^2}}.
\end{equation}
%
Following the treatment in \cite{Fatemiabhari:2025poq, Fatemiabhari:2025usn, Fatemiabhari:2026goj}, the proper coordinate takes the form
\begin{equation}\label{ysubspace}
    \dd y^2 =L^2_{\text{ABJM}} \left( r^2 \dd x_1^2 + \frac{\dd r^2}{r^2}\right).
\end{equation}
For which the Lagrangian is expressed as
\begin{equation}
    {\cal L}= \frac{k}{2} \sqrt{r^2-\frac{\dot{y}^2}{L_{\text{ABJM}}^2}}.
\end{equation}
From the above we obtain the associated proper momentum ${\cal P}_y$
\begin{equation}\label{Py_D0}
    |{\cal P}_y| = \frac{k}{2L_{\text{ABJM}}}\sqrt{\frac{r^4\dot{x}_1^2+\dot{r}^2}{r^4(1-\dot{x}_1^2)-\dot{r}^2}}=\frac{k}{2L_{\text{ABJM}}}\sqrt{\frac{{\cal P}_{x_1}^2}{{\cal H}^2 - {\cal P}_{x_1}^2}+\frac{4({\cal H}^2 - {\cal P}_{x_1}^2)}{k^2}t^2}\, .
\end{equation}
We follow \cite{Caputa:2024sux, Fan:2024iop, He:2024pox} and equate the rate of change of the complexity with this proper momentum, 
\begin{equation}\label{CdotD0}
    \dot{\cal C} \sim \frac{k}{2L_{\text{ABJM}}}\sqrt{\frac{{\cal P}_{x_1}^2}{{\cal H}^2 - {\cal P}_{x_1}^2}+\frac{4({\cal H}^2 - {\cal P}_{x_1}^2)}{k^2}t^2}\, .
\end{equation}
In the case of no-motion in the $x_1$-direction (${\cal P}_{x_1}=0$) we find
\begin{equation}
\dot{\mathcal{C}}\sim \frac{\cal H}{L_{\text{ABJM}}} ~t.\label{caputamomentum}  
\end{equation}
Equation (\ref{caputamomentum}) reproduces the result of \cite{Caputa:2024sux}, now using a genuine point-like excitation of the field theory (a D0 brane in the holographic description).
%
%
\subsection{A puzzle and a solution}
The result in eq.(\ref{CdotD0}) faces us with the following conundrum. It is clear that for  ${\cal P}_{x_1}\neq 0$, the presence of the conserved quantity changes the short times expansion of the rate-of-change of the complexity, respect to that in eq.(\ref{caputamomentum}). In fact, there is a cross-over time obtained from eq.(\ref{CdotD0})
\begin{equation}
    t_{\text{cross}}=\frac{k}{2}\frac{|{\cal P}_{x_1}|}{{\cal H}^2-{\cal P}_{x_1}^2},\label{tcross}
\end{equation}
separating the early-time behaviour  $\dot{{\cal C}}\sim \sqrt{\frac{{\cal P}_{x_1}^2}{{\cal H}^2 - {\cal P}_{x_1}^2} }$ from the long-time behaviour $\dot{{\cal C}}\sim \frac{2}{k}\sqrt{{({\cal H}^2 - {\cal P}_{x_1}^2)}}~~t$. The problem with this result is that it can be shown (see for instance \cite{Muck:2026top}) that the Krylov complexity ${\cal C}(t)$ must be an even function and that it must start at short times in a Taylor expansion $\dot{{\cal C}}\sim t$, in contradiction with the result in eq.(\ref{CdotD0}). One may attempt different answers to this dichotomy.

One possible radical solution is that the proposal of \cite{Caputa:2024sux,Fan:2024iop, He:2024pox} is invalid in the presence of conserved charges. This seems too extreme to us; as we do not have a better proposal at the moment,  we keep on working with the ideas presented in \cite{Caputa:2024sux}. Another possibility is that the definitions of the proper coordinate and proper  momentum in eqs.(\ref{ysubspace})-(\ref{Py_D0}) are not appropriate, but this definition works well in cases in which there is not an isometry and hence not a conserved charge, as is the case in quiver field theories \cite{Fatemiabhari:2025poq}. Here, we  propose a different solution. Our solution is based on working in a sector of {\it fixed conserved charge} ${\cal P}_{x_1}$. 

To implement this, we need to Legendre transform and work with the Routhian, instead of working with the Lagrangian in eq.(\ref{metric_and_lagrangian_D0}). In practice, we Legendre transform the Lagrangian in eq.(\ref{metric_and_lagrangian_D0}) along the isometric direction $x_1$, to obtain
\begin{eqnarray}
 & & {\cal P}_{x_1}=- \frac{k}{2}\frac{r^3\dot{x}_1}{\sqrt{r^4(1-\dot{x}_1^2)-\dot{r}^2}}\longrightarrow \dot{x}_1=-\frac{2 {\cal P}_{x_1}}{r^2}\sqrt{\frac{r^4-\dot{r}^2}{4 {\cal P}_{x_1}^2+ k^2 r^2}},\nonumber\\
 & & R={\cal P}_{x_1}\dot{x}_1 -{\cal L}= -\sqrt{g(r)\left(r^2-\frac{\dot{r}^2}{r^2}\right)}. 
 ~~~\text{Where}~~~ g(r)= \frac{4{\cal P}_{x_1}^2 + k^2 r^2}{4 r^2}.\label{routhian}
\end{eqnarray}
Working with the Routhian, we find an associated conserved Hamiltonian,
\begin{equation}
 {\cal H}= {\cal P}_r \dot{r} - R= r^3\sqrt{\frac{g(r)}{r^4-\dot{r}^2}} , ~~(\text{with}~~{\cal P}_r=\partial_{\dot{r}}R).   
\end{equation}
From here we recover eq.(\ref{eqdotr}) for $\dot{r}(t)$. In consequence, we find $r(t)$ in eq.(\ref{sols_x1_r}). We define a variable $\dot{y}^2= g(r)\frac{\dot{r}^2}{r^2}$ and the Routhian reads,
\begin{equation}
 R= -\sqrt{g(r) r^2-\dot{y}^2}.\label{routhianfinal}   
\end{equation}
The proper momentum  is calculated following the logic in eqs.(\ref{ysubspace})-(\ref{Py_D0}), but now applied to the Routhian in eq.(\ref{routhianfinal}). This gives $P_y=\sqrt{\frac{\dot{y}^2}{g(r) r^2-\dot{y}^2}}$. We associate this proper momentum with the rate-of-change of the complexity,
\begin{equation}
   \dot{{\cal C}}(t)\sim |P_y|= \sqrt{\frac{\dot{r}^2}{r^4-\dot{r}^2}}= \sqrt{\frac{4({\cal H}^2-{\cal P}_x^2)^3}{k^2 {\cal H}^4+ 4 {\cal P}_x^2 ({\cal H}^2-{\cal P}_x^2)^2 t^2}} ~~~t.\label{fullcomplexityD0}
\end{equation}
This expression satisfies that for short times $\dot{{\cal C}}\sim \frac{2\left( {\cal H}^2-{\cal P}_x^2\right)^{\frac{3}{2}}}{k {\cal H}^2}~ t$, resolving the contradiction pointed out below eq.(\ref{tcross}). It also recovers the result in eq.(\ref{caputamomentum}) for ${\cal P}_x=0$ (this requires a slight redefinition of $\dot{y}$ by  constant factor, or defining the Routhian multiplying it by a factor of $\frac{k}{L_{\text{ABJM}}}$). Interestingly, for large times it asymptotes to a constant value $\dot{C}\sim \frac{\sqrt{{\cal H}^2-{\cal P}_x^2}}{|{\cal P}_x|}$. This would not be the case if the conserved charge is associated with an isometry of $\mathbb{CP}^3$. We encounter some of these results when studying an extended object (in this work we focus on D3 branes) that fall in AdS$_5\times S^5$, see Section \ref{sectionD3branesinAdS}.
\subsection{A field theory calculation }
In this section we schematically indicate a calculation done in quantum field theory that reproduces the result for the Krylov spread complexity obtained by integrating eq.(\ref{caputamomentum}), namely ${\cal C}(t)=\frac{\cal H}{2L_{\text{ABJM}}}t^2$. It should be possible to extend this calculation to the case with a fixed conserved charge in eq.(\ref{fullcomplexityD0}). 

We propose to identify the falling D0 brane in AdS$_4$,  with a state created by a dressed\footnote{In ABJM, a monopole operator is not gauge invariant. We dress it with matter fields to cancel the electric charge and recover gauge invariance. See, for example \cite{Assel:2018wtj}} monopole operator \cite{Benna:2009xd}.  We consider the dressed monopole in \cite{Berenstein:2009sa}, with dimension $\Delta=\frac{k}{2}$, with $k$ being the Chern-Simons level of the ABJM theory. For a chiral primary (eigenstate of the Hamiltonian/dilatation operator), the Krylov chain contains only one site. What we  actually work with is the state created by a chiral primary ${\cal O}(x)$ after a time smearing (represented by the parameter $\epsilon$) has taken place, namely
\begin{equation}
|\psi \rangle = \frac{e^{- \epsilon H}}{\sqrt{{\cal N}}} {\cal O}(0)|0 \rangle . ~~~\text{with}~~{\cal N}= \langle 0|{\cal O}^\dagger(0)e^{-2\epsilon H} {\cal O}(0)|0 \rangle \label{state}
\end{equation}
This way of regulating gives the state a finite energy distribution and hence it can be decomposed as a  sum of infinite descendant states. This kind of regulations are common when considering a 'quench' in the CFT, see for example \cite{Caputa:2024sux}. 

We propose to identify  the survival amplitude
\begin{equation}
{A(t)=\langle \psi(0)|e^{- i H t}|\psi(0) \rangle }, \label{survivalt}   
\end{equation}
with the Euclidean correlation function
of two monopole operators,
\begin{equation}
G_E({x},{y})= \langle {\cal O}^{\dagger}(x){\cal O}(y) \rangle = \frac{C_{\Delta}}{|x-y|^{2\Delta}}.   
\end{equation}
$G_E(0,t)$ describes the propagator of the monopole  from zero-time to time $t$ (at position $\vec{x}=\vec{0}$) that gets destroyed by the anti-monopole at time $t$ and position $\vec{y}=\vec{0}$. After regulating to avoid the divergence at $t=0$ ---since the survival amplitude (\ref{survivalt}) must satisfy $A(0)=1$, we have
\begin{equation}
 A(t)=\frac{G_E(\vec{0},it+\beta)}{G_E(\vec{0},0,\beta)}=\Big[ \frac{\beta}{\beta +i t}\Big]^{\nu}, ~~\text{we define $\nu=2\Delta$}.\label{survivalutil}   
\end{equation}
Comparing with eq.(\ref{state}), we identify $\beta=2\epsilon$ (the smearing parameter). In what follows, we work with this expression, apply known formulas that give the momenta, the Lanczos coefficients and the wave functions in the Krylov basis. After this we calculate the spread complexity following \cite{Balasubramanian:2022tpr}. We describe the full calculation in \cite{ChHNRS}, here we limit ourselves to give the short time expansion of the complexity in terms of the momenta defined below. All the formulas we use (of which we just quote the results below), have been derived in \cite{Balasubramanian:2025xkj,Balasubramanian:2022tpr,Caputa:2025ozd,Caputa:2023vyr}.

Taylor expanding the survival amplitude (\ref{survivalutil}),
\begin{eqnarray}
& & A(t)=\Big[ \frac{\beta}{\beta +i t}\Big]^{\nu}=
\sum_{l=0}^\infty \frac{(-i t)^l}{l!}\mu_l,~~\text{the first few moments are}\label{momenta}\\
& & \mu_0=1,~~\mu_1=\frac{\nu}{\beta},~~
  \mu_2=\frac{\nu(\nu+1)}{\beta^2},~~\mu_3=\frac{\nu(\nu+1)(\nu+2)}{\beta^3},\nonumber\\
  & &\mu_4=\frac{\nu(\nu+1)(\nu+2)(\nu+3)}{\beta^4}.
  \label{eq:firstMoments}
\end{eqnarray}
Actually, it can be calculated  $\mu_l=\frac{\Gamma(\nu+l)}{\beta^l \Gamma(\nu)}$. All the Lanczos coefficients $a_n$ and $b_n$ can be written in terms of Hankel determinants, and can be calculated using a representation in terms of orthogonal polynomials. For the first few Lanczos coefficients we find,
\begin{eqnarray}
& &  a_0=\mu_1,
  ~~ b_1^2=\mu_2-\mu_1^2, \label{eq:firstab}\\
 & & a_1=\frac{\mu_3-2\mu_1\mu_2+\mu_1^3}{\mu_2-\mu_1^2},~~
  b_2^2=\frac{\Delta_2}{\Delta_1^2},
  \qquad
  \Delta_1=\mu_2-\mu_1^2,\nonumber\\
& &
  \Delta_2=\mu_2\mu_4-\mu_3^2-\mu_1^2\mu_4+2\mu_1\mu_2\mu_3-\mu_2^3.
  \label{eq:Delta2explicit}
\end{eqnarray}
All other Lanczos coefficients can be written in terms of the momenta $\mu_l$ and the Hankel determinants of higher order. See \cite{Balasubramanian:2025xkj, Balasubramanian:2022tpr, Caputa:2025ozd, Caputa:2023vyr} for explicit formulas. The complexity in a short time expansion is
\begin{eqnarray}
& & {\cal C}(t)=b_1^2t^2+b_1^2\left[\frac{b_2^2}{6}-\frac{b_1^2}{3}-\frac{(a_1-a_0)^2}{12}\right]t^4+{\cal O}(t^6),\nonumber\\
& & \text{or, in terms of the first moments,}\nonumber\\
& & {\cal C}(t)=(\mu_2-\mu_1^2)t^2+
  \left[
  \frac{\Delta_2}{6\Delta_1}-\frac{\Delta_1^2}{3}
  -\frac{(\mu_3-3\mu_1\mu_2+2\mu_1^3)^2}{12\Delta_1}
  \right]t^4+{\cal O}(t^6).\label{complexitymuss}
\end{eqnarray}
Computing explicitly, we find
\begin{equation}
 {\cal C}(t)= \frac{\nu}{\beta^2}~t^2= \frac{2\Delta}{\beta^2}~t^2.   
\end{equation}
In this case there are no higher order corrections, all coefficients of $t^{2n}$ with $n>1$ vanish.  In retrospective, this is a consequence that the survival amplitude is that of a zero temperature CFT. This coincides with the result in eq.(\ref{caputamomentum}) after identifying CFT parameters $(\beta,\Delta)$ with holographic side parameters $({\cal H},L_{\text{ABJM}})$, this suggests the identification $\Delta\sim \frac{\cal H}{L_{\text{ABJM}}}$.

We study now fluctuations of our probe D0 and their impact on the complexity.

\subsection{Fluctuations along the radial direction}
We are probing the ABJM background with a genuine excitation of the theory. In fact the D0 brane is in the spectrum of IIA and corresponds to a monopole in the ABJM CFT$_3$. This D0 probe (or monopole) also carries internal degrees of freedom, encoded by the BI–WZ action. In this section we study the effect of these internal degrees of freedom on the complexity.

We consider small fluctuations along the radial direction, for the classical solution given by eq.\eqref{sols_x1_r}, and their influence in the complexity. This is a consistent truncation of the full fluctuated system. We make a further simplification by choosing ${\cal P}_{x_1}=0$, in order to illustrate the effect of the fluctuations in the simplest possible scenario: the falling "particle" in $\text{AdS}$,  calculating corrections to the result of \cite{Caputa:2024sux}. Setting
\begin{equation}
    x_1 = 0\quad \& \quad r=r_{\cl}(t)+\varepsilon \delta r(t),
\end{equation}
where $|\varepsilon|\ll 1$ and $r_\cl$ is given by

\begin{equation}\label{eq:r_classical}
    r_\cl (t) = \frac{2{\cal H}}{\sqrt{k^2 +4{\cal H}^2 t^2}},
\end{equation}
we can expand up to second order to find the solution for $\delta r(t)$ by reading the ${\cal O}(\varepsilon^2)$ Lagrangian

\begin{equation}
    {\cal L}_{\delta r}^{(2)}=\frac{k}{2(r_\cl^4-\dot{r}_\cl^2)^{3/2}}\Big[ \frac{\dot{r}_\cl^2(\dot{r}^2_\cl-3r_\cl^4)}{r_\cl^3}\delta r^2 - \frac{\dot{r}_\cl(\dot{r}_\cl^2-3r_\cl^4)}{r_\cl^2}\delta r\delta \dot{r}-\frac{r_\cl^3}{2}\delta \dot{r}^2 \Big] .
\end{equation}

\begin{figure}[t]
    
        \centering
        \includegraphics[width=0.6\textwidth]{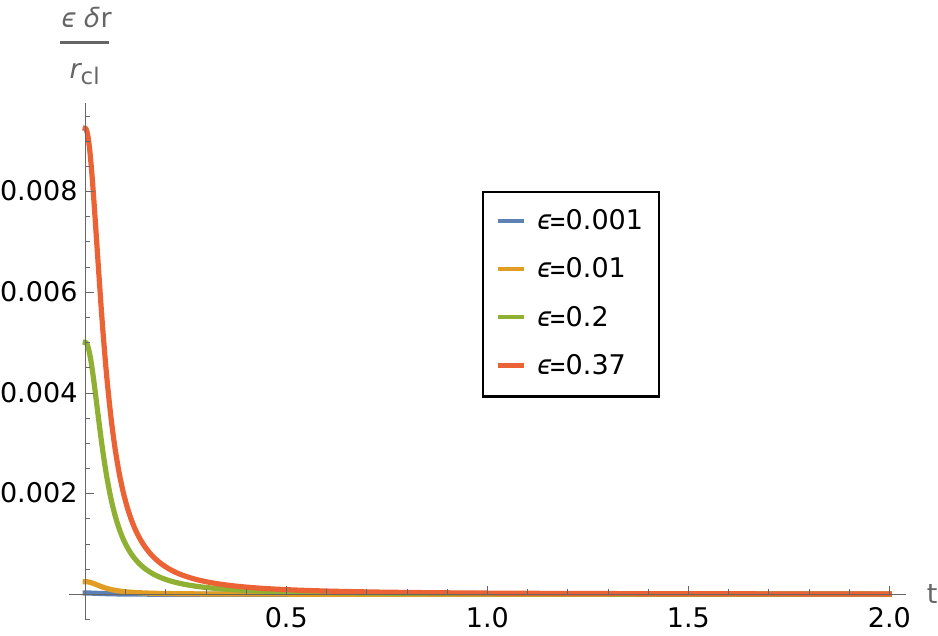}
    
    \caption{Plot of the ratio $\frac{\varepsilon ~\delta r}{r_\cl}$. The following values have been chosen for the parameters: $L_{\text{ABJM}}=k=1,c_1=0.5,{\cal H}=10$.}
\label{fig:flucratio}
\end{figure}

The solution for the fluctuation, which we plot over $r_\cl$ in Figure \ref{fig:flucratio}, is found by solving its equations of motion and reads
\begin{equation}\label{sol_fluc_dr}
    \delta r(t)= \frac{c_1}{(k^2+4{\cal H}^2 t^2)^{3/2}}+\frac{c_2 t}{(k^2+4{\cal H}^2 t^2)^{3/2}}.
\end{equation}
Here, $c_1$ and $c_2$ are arbitrary constants. The conditions
\begin{equation}
    \left| \frac{\varepsilon\delta r}{r_\cl} \right|\ll 1,\quad \left| \frac{\varepsilon\delta \dot{r}}{\dot{r}_\cl}\right|\ll 1,
\end{equation}
force us to set $c_2=0$ and we can combine $c_1$ and $\varepsilon$ into a single perturbative parameter $\tilde{\lambda} = \varepsilon c_1$ with $|\tilde{\lambda}|\ll 1$. This will allow us to use the solution of the linearised problem \eqref{sol_fluc_dr}.

To calculate the Krylov complexity we need to introduce the proper coordinate, defined as

\begin{equation}
    \dd y^2 = L^2_{\text{ABJM}} \frac{\dd r^2}{r^2} ,
\end{equation}
in terms of which the zeroth order Lagrangian \eqref{metric_and_lagrangian_D0} takes the usual form

\begin{equation}
    {\cal L}^{(0)}=\frac{k}{2}\sqrt{r^2-L^{-2}_{\text{ABJM}} \dot{y}^2},
\end{equation}
and the proper momentum is given by
\begin{equation}\label{Py_generic}
    {\cal P}_y = - \frac{k}{2L^2_{\text{ABJM}}}\frac{\dot{y}}{\sqrt{r^2 -  L^{-2}_{\text{ABJM}}\dot{y}^2}}.
\end{equation}

The idea is to study the dynamical influence of the fluctuations at the level of \eqref{Py_generic}, by including them in the definition of the proper coordinate in ${\cal L}^{(0)}$ and without expanding. We thus propose

\begin{equation}
    \dot{y} = L_{\text{ABJM}} \frac{\dot{r}_{\cl}+\varepsilon\delta \dot{r}}{r_{\cl}+\varepsilon\delta r}.
\end{equation}

This results in the following expression for the rate of change of complexity in the presence of small fluctuations

\begin{equation}\label{Cdot_fluc}
   \dot{\cal C}_{\text{fluc}}\sim|P_y|=\frac{k|\dot{r}_{\text{cl}} +\varepsilon~\delta \dot{r}|}{2L_{\text{ABJM}}\sqrt{(r_{\text{cl}} +\varepsilon ~\delta r)^4 -(\dot{r}_{\text{cl}}+\varepsilon~ \delta\dot{r})^2}}\, .
   %
\end{equation}
The above formula is {\it exact} and no expansion has been made. In principle one could solve numerically the non-linear problem and obtain the solution for the fluctuations (including higher order terms in the Lagrangian), which will enter in the above. We choose to study analytically the perturbative corrections in $\tilde{\lambda}=\varepsilon c_1$ and therefore substituting \eqref{eq:r_classical} and \eqref{sol_fluc_dr} in the above and expanding yields

\begin{equation}\label{eq:Cdot_fluctuated}
    \dot{\cal C}_{\text{fluc}}\sim \frac{\mathcal{H} t}{ L_{\text{ABJM}}} +\frac{t}{2k^2 L_{\text{ABJM}}}\tilde{\lambda}-\frac{3k^2t - 6{\cal H}^2t^3}{4k^4{\cal H}L_{\text{ABJM}}(k^2+4{\cal H}^2 t^2)}\tilde{\lambda}^2
+{\cal O}(\tilde{\lambda}^3).
\end{equation}
The leading correction due to the perturbation is correcting the Hamiltonian of the falling D0, 
\begin{equation}\label{eq:Cdot_first_order_corr}
    \dot{\cal C}_{\text{fluc}}^{(1)}= \frac{{\cal H}_{\text{fluc}}^{(1)}}{L_{\text{ABJM}}}t,\quad  {\cal H}_{\text{fluc}}^{(1)}={\cal H} +\frac{\tilde{\lambda}}{2k^2}.
\end{equation}
For large times, the ${\cal O}(\tilde{\lambda}^2)$ corrected expression of the Hamiltonian is
\begin{equation}
    \dot{\cal C}_{\text{fluc}}^{(2)}\approx \frac{{\cal H}_{\text{fluc}}^{(2)}}{L_{\text{ABJM}}}t,\quad {\cal H}_{\text{fluc}}^{(2)}={\cal H}+\frac{\tilde{\lambda}}{2k^2}+\frac{3\tilde{\lambda}^2 }{8k^4{\cal H}}.
\end{equation}
{{\bf An interpretation}: Let us present another way of thinking about the above calculation, which actually showcases that the fluctuation presented here does {\it not} express a dynamical excitation of the probe D0. For this one has to include fluctuations in the internal space, which we leave for future work. One may rewrite the classical solution \eqref{eq:r_classical} in terms of the initial radial point $r_{\text{UV}}$ of the trajectory, which is subject to the condition $\dot{r}(t=0)=0$. Using \eqref{eqdotr} (in the case where ${\cal P}_{x_1}=0$) we have

\begin{equation}
    r_\cl (t;r_\UV) = \frac{r_{\text{UV}}}{\sqrt{1+r_{\text{UV}}^2 t^2}},\qquad r_{\text{UV}}=\frac{2{\cal H}}{k}.
\end{equation}

We can now consider a small deviation from the parameter $r_\UV$ as $r_\UV\mapsto r_\UV + \varepsilon \delta r_\UV$, leading to the expansion

\begin{equation}
    r_\cl(t;r_\UV+\varepsilon \delta r_\UV)=r_\cl (t) + \frac{\varepsilon \delta r_\UV}{(1+r_\UV^2 t^2 )^{3/2}}+ {\cal O}(\delta r_\UV^2),
\end{equation}
from which we can identify up to first order
\begin{equation}\label{eq:delta_r_cl}
    \delta r_\cl (t;r_\UV)\equiv r_\cl(t;r_\UV + \varepsilon \delta r_\UV)-r_\cl (t;r_\UV)=\frac{\delta r_\UV }{(1+ r_\UV^2t^2)^{3/2}},
\end{equation}
which is exactly the solution we found in \eqref{sol_fluc_dr} (for $c_2=0$) after identifying\footnote{we should have agreement of \eqref{sol_fluc_dr} and \eqref{eq:delta_r_cl} at $t=0$.} $\delta r_\UV = c_1/k^3$ In this sense the fluctuation we calculated is a rescaled version of the classical trajectory. However, this expresses the response of the system to a change in its initial condition, and given that $r_\UV$ is related to the Hamiltonian of the trajectory, we have the following change 
\begin{equation}
    \delta {\cal H} = \frac{k}{2}\delta r_\UV=\frac{\tilde{\lambda}}{2k^2},
\end{equation}
agreeing with \eqref{eq:Cdot_first_order_corr}. One may then interpret the corrections to the complexity in \eqref{eq:Cdot_fluctuated} as studying how it responds to small changes in the prepared initial state which is chosen to calculate it in field theory \cite{Caputa:2024sux}.}

Let us summarise. The calculation of \cite{Caputa:2024sux} considers a {\it rigid particle} in AdS$_3$. Here, we discussed  a particle in AdS$_4$. {In our case, the probe particle belongs to the spectrum  of the IIA string, has an inherent dynamics for its internal excitations, which influence  the complexity, as we found.}

The analysis of the D0 brane therefore provides a useful  top-down realisation of the proper-momentum proposal for Krylov spread complexity. Starting from a genuine type IIA excitation in the ABJM background, we reproduced the expected falling-particle result in the absence of conserved spatial momentum, identified the apparent conflict produced by a non-zero conserved charge, and resolved it by working at fixed charge through the Routhian. In this fixed-charge description the short-time behaviour of the rate of complexity is again linear in time, as required by the general structure of Krylov complexity, while the uncharged limit smoothly reproduces the known result. The same D0 system also admits a field-theory interpretation in terms of a regulated monopole two-point function, and its radial fluctuations give controlled corrections to the effective energy entering the complexity growth. These observations suggest that the Routhian prescription is not merely a technical device for pointlike probes, but a natural way of treating holographic spread complexity in sectors with conserved charges. We now test this idea in a more elaborate setting, replacing the D0 brane by an extended non-BPS D3 brane falling in $AdS_5\times S^5$ and carrying angular momentum on the internal sphere.


\section{D3 branes in AdS$_5\times S^5$}\label{sectionD3branesinAdS}
In this section we consider the AdS$_5\times S^5$ background in Type IIB. We study a D3 brane that falls in AdS$_5$ and rotates in $S^5$. The rotation makes the brane non-BPS.

The type IIB  background  contains a metric and four-form Ramond gauge field, with all other fields vanishing. The string frame background is,
\begin{eqnarray}
& &\dd s^2   =\frac{r^2}{L^2}\left( -\dd t^2+ \dd x_1^2+ \dd x_2^2+\dd x_3^2\right)+\frac{L^2}{r^2}\dd r^2+ L^2 \dd \Omega_5^2,\nonumber\\
& & \dd\Omega_5^2=\dd \theta_1^2+\sin^2\theta_1 \dd\theta_2^2+\sin^2\theta_1\sin^2\theta_2 \dd \theta_3^2 + \sin^2\theta_1\sin^2\theta_2 \sin^2\theta_3 \dd \theta_4^2 + \sin^2\theta_1\sin^2\theta_2 \sin^2\theta_3 \sin^2\theta_4 \dd\psi^2,\nonumber\\
& &C_4=\frac{r^4}{L^4}\dd t\wedge \dd x_1\wedge \dd x_2\wedge \dd x_3.\label{IIBbackground} 
\end{eqnarray}
We consider a probe D3-brane  extended on $[t,x_1,x_2,x_3]$ that falls in AdS$_5$ according to $r(t)$ and also rotates inside the equator of the $S^5$, with $\psi(t)$, for fixed values of the angles $\theta_i=\theta_{0,i}$.
The Born-Infeld-Wess-Zumino action of this brane is
\begin{equation}\label{L_brane_angularmomentum_generic}
    \mathcal{L} = \nu \left(\frac{r^{3}}{L^{3}}\sqrt{\frac{r^{2}}{L^{2}}-\frac{L^{2}\dot{r}^{2}}{r^{2}}-\lambda^{2} \dot{\psi}^{2}}- \frac{r^{4}}{L^{4}}\right).
\end{equation}
 We have defined the parameters 
 \begin{equation}
     \lambda^2= L^2\sin^2\theta_{1,0}\sin^2\theta_{2,0} \sin^2\theta_{3,0} \sin^2\theta_{4,0}, ~~\text{and}~~  \nu=T_{D3}\int dx_1 dx_2 dx_3,~ \text{with}~ (2\pi)^3 g_s \alpha'^2 T_{D3}=1.\nonumber\end{equation} 
In particular, a solution exists to the equations of motion for $\theta_i(t)=\frac{\pi}{2}$, in which case $\lambda=L$. Other solutions exist, but they are probably unstable, and we do not discuss them here.

The Hamiltonian is given by
\begin{equation}
    H = P_{r}\dot{r}+P_{\psi}\dot{\psi}-\mathcal{L} = \nu\frac{r^{4}}{L^4}\left( 1-\frac{r}{L\Delta}\right),
~~\text{with}~~\Delta = \sqrt{\frac{r^{2}}{L^{2}}-\frac{L^{2}\dot{r}^{2}}{r^{2}}-\lambda^{2} \dot{\psi}^{2}}.    
    \label{hamiltonianded3}
\end{equation}
We consider the case in which the brane falls from a position $r_{\text{UV}}$ with zero initial velocity $\dot{r}(0)=0$ and  nonzero initial angular velocity $\dot{\psi}_0\neq 0$. The  Hamiltonian is 
\begin{equation}
    H = \nu\frac{r_{\text{UV}}^{4}}{L^{4}}\left(1-\frac{r_{\text{UV}}}{L\sqrt{\frac{r_{\text{UV}}^{2}}{L^{2}}-\lambda^{2} \dot{\psi}_{0}^{2}}}  \right) \, .
\end{equation}
The Hamiltonian is non-zero provided that $\dot{\psi} \neq 0$. In fact, for $\dot{\psi}=0$ the probe is BPS and does not move (or moves with constant $\dot{r}(t)=v_0$). The energy of the probe is $\mathcal{H} = -H$.\\ The units of the quantities characterising the motion of D3 probes are,
\begin{equation*}
[\lambda]=[L]=[r]=[t]=[x_i]=\text{length}, ~~[H]=[\nu]=[{\cal L}]=\frac{1}{\text{length}},~[\theta_i]=[\psi]=1.    
\end{equation*}
One may consider similar probes on different backgrounds of the form AdS$_{p+2}\times \Sigma^{8-p}$ with electric Ramond potential. In fact, D1 (or F1) probing the D1/D5 (or NS5/F1) system, D2  branes in ABJM, D4 branes in AdS$_6\times \Sigma^4$ display similar Lagrangians as that in eq.(\ref{L_brane_angularmomentum_generic}) and the dynamics generalises the study we present below. See \cite{ChHNRS} for details.

From the Lagrangian in eq.(\ref{L_brane_angularmomentum_generic}) we can obtain the conserved quantity $P_{\psi} \equiv -{\cal J}$ associated with translations in $\psi$. Using this and the Hamiltonian in eq.(\ref{hamiltonianded3}), one finds  first order equations (solving the Euler-Lagrange equations) for $r(t)$ and $\psi(t)$. In fact,
\begin{eqnarray}
& &    \dot{\psi} = \frac{L^{2} {\cal J} r^{2} }{\lambda^{2}(L^{4}\mathcal{H}+\nu r^{4})} \, ,\\
& &   \dot{r} =- \frac{r^{2}}{L^{2}}\frac{\sqrt{L^{4}\mathcal{H}(L^{4}\mathcal{H}+2\nu r^{4})-\lambda^{-2}L^{6}{\cal J}^{2}r^{2}}}{L^{4}\mathcal{H}+\nu r^{4}} \, .
    \label{rdotj}
\end{eqnarray}
Let us discuss this dynamical system in more detail.
Consider the square of equation \eqref{rdotj}
\begin{equation}
    \dot{r}^{2} = \underbrace{\frac{r^{4}}{L^{4}}\frac{1}{(L^{4}\mathcal{H}+\nu r^{4})^{2}}}_{A(r)} \, \underbrace{(L^{4}\mathcal{H}(L^{4}\mathcal{H}+2\nu r^{4})-\lambda^{-2}L^{6}{\cal J}^{2}r^{2})}_{F(r)} \, .\label{julian}
\end{equation}
For $t= 0$ we have $r = r_{\text{UV}}$, $\dot{r} = 0$. Since $A(r)>0$,  $F(r_{\text{UV}}) = 0$ which implies the relation
\begin{equation} (\mathcal{H}L^{4})^{2}+2\nu\mathcal{H}L^{4}r_{\text{UV}}^{4} = \frac{{\cal J}^{2}L^{6}}{\lambda^{2}}r_{\text{UV}}^{2} \, .
    \label{Fcond}
\end{equation}
Expanding equation \eqref{rdotj} near $r_{\text{UV}}$, we get
\begin{equation}
    \dot{r}^{2} = A(r_{\text{UV}})F'(r_{\text{UV}})(r-r_{\text{UV}})+\mathcal{O}((r-r_{\text{UV}})^{2}) \, ,
\end{equation}
we used the fact that $F(r_{\text{UV}}) = 0$. Notice that, since the probe is falling, then $r(t)\leq r_{\text{UV}}$ and given that $A(r)$ is positive, then $F'(r_{\text{UV}})< 0$. Using eq. \eqref{Fcond} leads to
\begin{equation}
    \mathcal{H} > 2\nu\left(\frac{r_{\text{UV}}}{L}  \right)^{4} \, .\label{constr1}
\end{equation}
The latter expression gives a lower bound to the energy for which the probe falls. Using equation \eqref{Fcond} we can rewrite the inequality in terms of the angular momentum as 
\begin{equation}
    \frac{{\cal J}^{2}}{\lambda^{2}\mathcal{H}^{2}} < 2\left(\frac{L}{r_{\text{UV}}}\right)^{2}.
    \label{jcond}
\end{equation}
The value of ${\cal J}$ must  satisfy eq.\eqref{jcond}, otherwise, the probe does not fall. In fact, for values of ${\cal J}$ not satisfying this bound, a centrifugal barrier arises that prevents the D3 to fall in the radial direction. Let us clarify this. For fixed $r_{\text{UV}}$ and positive energy ${\cal H}$, the zero-velocity initial condition fixes ${\cal J}$ through eq.(\ref{Fcond}). The particle falls inward only if $F'(r_{\text{UV}})<0$, equivalently eq.(\ref{constr1}). In terms of ${\cal J}$, this becomes eq.(\ref{jcond}). If this is violated, the position $r_{\text{UV}}$ is not the upper end of a falling trajectory, the effective potential develops a centrifugal obstruction described below.

\begin{figure}[t]
    \centering
    \includegraphics[width=0.6\linewidth]{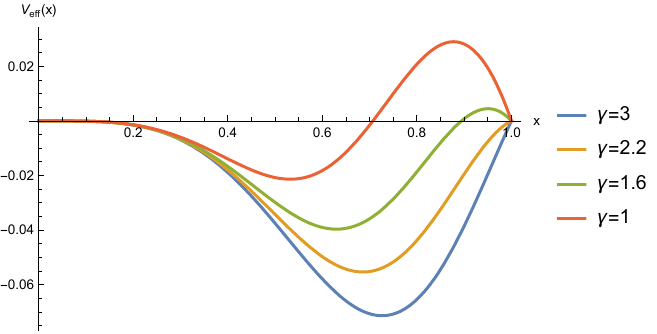}
    \caption{Plot of $V_{\text{eff}}$ with respect to $x$ for a range of $\gamma$ values. For $\gamma<2$, a centrifugal barrier prevents the probe to fall towards $x<1$. For $\gamma>2$ the probe falls.}
    \label{fig:Veffx}
\end{figure}

More quantitatively the content above can be described as follows. Using  dimensionless variables 
\begin{equation}
x(t)=\frac{r(t)}{r_{\text{UV}}},~~\gamma=\frac{\cal H}{\nu}\left(\frac{L}{r_{\text{UV}}}\right)^4,~~\beta=\left(\frac{{\cal J} r_{\text{UV}}}{\lambda {\cal H}L}\right)^2,~~\tau=\frac{r_{\text{UV}}t}{L^2},   \label{adim-variables} 
\end{equation}
equation (\ref{julian}) reads,
\begin{eqnarray}
& & \left(\frac{\dd x}{\dd\tau}\right)^2+V_{\text{eff}}(x)=0,
~~~~V_{\text{eff}}(x)=  \frac{x^4}{(x^4+\gamma)^2} \left[ \gamma(\gamma+2) x^2 -\gamma^2 -2\gamma x^4\right]. \label{julian2} 
\end{eqnarray}
We have used the constraints in eqs.(\ref{Fcond})-(\ref{jcond}), that in these variables read $\beta=\frac{\gamma+2}{\gamma}$ and $\gamma>2$. Equation (\ref{julian2}) corresponds with the dynamics of a particle with zero energy in the potential $V_{\text{eff}}$. If one takes $\gamma<2$, this effective potential develops a centrifugal barrier that prevents motion below $x=1$. See Figure \ref{fig:Veffx}.

On the other hand, equation \eqref{rdotj} in these variables reads,
\begin{equation}
\dot{x}=-\frac{x^2}{\gamma+x^4}\sqrt{\gamma(\gamma+2 x^4)-\gamma(\gamma+2) x^2}. \label{lio}   
\end{equation}

An exact solution to the differential equation (\ref{lio}) can be found in terms of incomplete elliptic integrals. After conveniently choosing an integration constant such that $x(\tau=0)=1$, the result in terms of the Elliptic functions of the first kind (denoted EllipticF) and second kind (denoted EllipticE) reads,
\begin{eqnarray}
& & \tau=-\int dx \frac{x^4+\gamma}{x^2\sqrt{-\gamma(\gamma+2)x^2 +\gamma^2+2\gamma x^4}} \nonumber\\
& & =\frac{\sqrt{\gamma(\gamma-2 x^2)(1-x^2)}}{\gamma~ x}+\frac{3}{2}\left(\text{EllipticE}[\arcsin x,\frac{2}{\gamma}] -\text{EllipticF}[\arcsin x,\frac{2}{\gamma}] \right)   + \tau_0.\label{catter}
\end{eqnarray}

The value of $\tau_0$ is chosen such that at $x=1$ we find $\tau=0$.
This expression (\ref{catter}) is not very convenient to work with.
In contrast, it is useful to write a series expansion of the solution to the differential equation (\ref{julian2}). In terms of the dimensionless variables defined in eq.(\ref{adim-variables}), we find
\begin{eqnarray}
    & & x(\tau)\approx 1- (\gamma-2)\Big[ \frac{ \gamma~\tau^2}{2(1+\gamma)^2}+ {\cal O}(\tau^4)\Big],~~\text{for}~~\tau\to 0,\label{expansionshort}\\
    & & x(\tau)\approx \frac{1}{\tau} -\frac{(\gamma+2)}{2~\gamma~\tau^3} +\frac{3(\gamma+2)^2}{8\gamma^2~\tau^5}
    +{\cal O}\left(\frac{1}{\tau^7}\right), ~~\text{for}~~\tau\to\infty.\label{expansionlong}
\end{eqnarray}

Let us now study the spread complexity associated with this falling D3 brane.
\subsection{Krylov spread complexity}
To (holographically) calculate the Krylov spread complexity \cite{Balasubramanian:2022tpr}, we  first observe that there is a conserved charge. This conserved charge prevents us from having completely static initial conditions. Indeed, whilst $\dot{r}(0)=0$ is attainable, the presence of $\dot{\psi}(0)\neq 0$ (which makes the brane mass/tension bigger than the brane charge) is needed for the brane to fall. Hence, zero initial velocities are not attainable. We are in a situation similar to that in Section \ref{SectionABJMm}, where we explained that we must work with a constant value of the conserved quantity. To work with a fixed value of angular momentum we repeat the treatment in Section \ref{SectionABJMm}, adapted to the present case. Schematically, the procedure goes as follows:
\\
First, we define the Routhian. To do this, we start with the Lagrangian in eq.(\ref{L_brane_angularmomentum_generic}), we calculate the conserved momentum ${\cal J}=-P_\psi=-\frac{\delta {\cal L}}{\delta\dot{\psi}}$. Using this we find $\dot\psi$ as function of $r,\dot{r}$ and ${\cal J}$. This gives,
\begin{equation}
    \dot{\psi }=  \frac{L^2{\cal J}}{\lambda ~ r}\sqrt{\frac{r^4 - L^4 \dot{r}^2}{L^{6}{\cal J}^2 + \lambda^2 \nu^2 r^{6}}}.
\end{equation}
Then, we calculate the Routhian for this system 
\begin{equation}\label{ruthian_brane}
    {\cal R}_{\text{brane}} =P_\psi \dot{\psi}-{\cal L}= -{\cal J}\dot{\psi}- {\cal L} = \nu \frac{r^{4}}{L^{4}} - \sqrt{\frac{({\cal J}^2 L^{6} + \lambda^2 \nu^2 r^{6})}{\lambda^2 L^{6}}\left(\frac{r^2}{L^2} - \frac{L^2 \dot{r}^2}{r^2}\right)}.
\end{equation}
Note that the   kinetic part is of the form in eq.\eqref{routhian}, with $g(r)=\frac{({\cal J}^2 L^{6} + \lambda^2 \nu^2 r^{6})}{\lambda^2 L^{6}}$.  Using the Routhian as the new Lagrangian, the corresponding conserved Hamiltonian reads,
\begin{equation}
    {\cal H}_{\text{routh,b}}=-\nu\frac{r^{4}}{L^{4}}+\frac{r^3}{L^{4}\lambda}\sqrt{\frac{L^{6}{\cal J}^2+\lambda^2\nu^2 r^{6}}{r^4 -L^4 \dot{r}^2}}.
\end{equation}
With this, we find $\dot{r}$
\begin{equation}
    \dot{r} = - \frac{r^2}{L^2} \frac{\sqrt{  \lambda^2  {\cal H}_{\text{routh,b}}^2L^{8}-{\cal J}^2 L^{6}r^2 + 2    {\cal H}_{\text{routh,b}} \lambda ^2\nu L^{4} r^{4}}}{\lambda(    {\cal H}_{\text{routh,b}} L^{4} + \nu r^{4})}.
\end{equation}
 Note that this is the result in eq.(\ref{rdotj}).
The relation in eq.(\ref{Fcond}) follows, and the analysis of the motion in eqs.(\ref{julian})-(\ref{julian2}) is also valid for the Routhian dynamics.
%
With the Routhian in eq.(\ref{ruthian_brane}), we define a {\it proper coordinate}
\begin{equation}
 \dot{y}^2=   \left(\frac{{\cal J}^2 L^{6} + \lambda^2 \nu^2 r^{6}}{\lambda^2 L^{6}}\right)\left( \frac{L^2 \dot{r}^2}{r^2}\right)= g(r) \left(\frac{L^2\dot{r}^2}{r^2}\right).
\end{equation}
From this we implicitly write  $r(y)$, which allows us to rewrite the Routhian and the associated proper momentum as,
\begin{eqnarray}
& & {\cal R}_{\text{brane}} = \nu \frac{r^{4}}{L^{4}} - \sqrt{\frac{({\cal J}^2 L^{6} + \lambda^2 \nu^2 r^{6})}{\lambda^2 L^{6}}\left(\frac{r^2}{L^2} \right)-\dot{y}^2},\nonumber\\
& &P_y= \frac{\delta {\cal R}_{\text{brane}}}{\delta \dot{y}}=\frac{\dot{y}}{\sqrt{\frac{({\cal J}^2 L^{6} + \lambda^2 \nu^2 r^{6})}{\lambda^2 L^{6}}\left(\frac{r^2}{L^2} \right)-\dot{y}^2}}.\label{PyD3}
\end{eqnarray}
Notice that the units of $[\dot{y}]=\frac{1}{\text{length}}$ and that $P_y$ is dimensionless. It is instructive to write the expression for $P_y$---that we equate with the rate-of-change of the complexity respect to the dimensionless time $\tau$---in terms of  the dimensionless variables defined in eq.(\ref{adim-variables}). Using  $\gamma\beta=(\gamma+2)$, we have
\begin{eqnarray}
& & \left(\frac{{\cal J}^2}{\lambda^2}+ \nu^2\frac{r^6}{L^6} \right)\to \left(\frac{r_{\text{UV}}}{L}\right)^6\nu^2\left[\gamma(\gamma+2)+x^6 \right],~~~\frac{r}{L}\to \frac{r_{\text{UV}}}{L} ~x,\nonumber\\
& & \dot{y}^2\to \left( \frac{r_{\text{UV}}}{L}\right)^8 \nu^2 \left[\gamma (\gamma +2)+x^6 \right]\left(\frac{\dot{x}}{x}\right)^2.\label{changeadim} 
\end{eqnarray}
The proper momentum in eq.(\ref{PyD3}) reads,
\begin{equation}
 P_y=\frac{\dot{x}(\tau)}{\sqrt{x^4(\tau)-\dot{x}^2(\tau)}}.\label{propermomentumadim} 
\end{equation}
We have denoted by $\dot{x}$ the derivative with respect to the dimensionless time $\tau$. We follow the proposal of  \cite{Caputa:2024sux, Fan:2024iop,He:2024pox}, now in terms of the proper momentum $P_y$ above, setting $\dot{C}\sim |P_y|$.
Using  the early  and late times expansions in eqs. (\ref{expansionshort}) and (\ref{expansionlong}), we find
\begin{eqnarray}
& & \dot{{\cal C}}\sim (\gamma-2)\Bigg[ \frac{\gamma}{(\gamma+1)^2}~\tau + 
{\cal O}(\tau^3)\Bigg],~\text{for}~\tau\to 0\label{complexityshorttau},\\
& & \dot{{\cal C}}\sim \sqrt{\frac{\gamma}{\gamma+2}}~\tau + 
{\cal O}(\frac{1}{\tau^3}),~~\text{for}~\tau\to\infty. \label{complexitylongtau}
\end{eqnarray}
We can go back to the original (physical) variables by changing $\tau\to \frac{r_{\text{UV}} }{L^2} t$ and $\gamma= \frac{{\cal H}_{\text{routh,b}}}{\nu}\left(\frac{L}{r_{\text{UV}}}\right)^4$. Also, in eqs.(\ref{complexityshorttau})-(\ref{complexitylongtau}), the derivatives of the complexity are with respect to the dimensionless time $\tau$. We should also be mindful of the constraint $\gamma\beta=(\gamma+2)$, that is equivalent to eq.(\ref{Fcond}).
The expression in eq.(\ref{complexityshorttau}) has the required linear behaviour and depends on the various parameters as well as the conserved charge ${\cal J}$. On the other hand, for long times, we find a linear behaviour for the rate-of-change of the spread complexity, with subleading corrections. It is nice to observe that in the case $\gamma=2$ (the brane does not fall), the complexity is constant. In conclusion, the spread complexity associated with an initial state created by an extended operator, in this case represented by our non-BPS D3 brane, is {\it not equal} to the spread complexity of a massive, uncharged and unexcited  state.

The D3-brane example confirms and extends the lesson of the D0-brane analysis. The angular velocity on $S^5$ makes the probe non-BPS and gives rise to a conserved angular momentum, so that the falling motion must again be described in a fixed-charge sector. The Routhian treatment leads to a simple proper momentum, $P_y=\dot x/\sqrt{x^4-\dot x^2}$, in terms of the dimensionless radial variable, and the resulting rate of spread complexity has the expected short-time behaviour $\dot C\sim \tau$, with a coefficient controlled by the energy, angular momentum and initial radial position of the brane. The same analysis also reveals a sharp physical distinction between falling and non-falling trajectories: when the bound on the angular momentum is violated, a centrifugal barrier prevents the brane from moving inward, while the limiting case $\gamma=2$ gives a static configuration with constant complexity. Thus, the spread complexity associated with an extended non-BPS operator is sensitive not only to the energy of the state, but also to its conserved internal charges and to the possibility of radial infall.
It would be interesting to repeat this type of calculations in different AdS-spaces with different brane probes, for example those in 
\cite{Gaiotto:2009gz, Lozano:2019zvg, Lozano:2020sae, Legramandi:2021uds, Akhond:2021ffz, Assel:2011xz, Macpherson:2024frt, Anabalon:2026yxk}. In these cases, we anticipate an interesting interplay between the conserved charge and the motion of the brane within the quiver SCFT.

In the next section we study a complementary extended excitation, a fundamental string wrapping directions inside $S^5$ and falling in $AdS_5$. This example will show how a wound string can reduce to an effective massive particle, and it  clarifies the distinction between Noether charges that require a fixed-charge Routhian treatment and winding data that can be incorporated directly into the effective mass of the falling object.

\section{Strings in AdS$_5\times S^5$}\label{seccion-4}
%
In this section we study the spread complexity in ${\cal N}=4$ SYM. We follow the proposal of  \cite{Caputa:2024sux, Fan:2024iop, He:2024pox}. Along the same lines of the study in Section \ref{SectionABJMm}, we consider a genuine excitation of the field theory, represented holographically, by an F1 string that falls along the radial coordinate and wraps the internal directions of AdS$_5\times S^5$. The calculation with a D1 brane is similar.
In the following $(\tau,\sigma)$ denote the worldsheet coordinates of the fundamental string.

The  parametrisation of $AdS_5 \times S^5$ spacetime chosen in this case is,
\begin{align}
    \mathrm{d}s^2 =& -\frac{r^2}{L^2} \mathrm{d}t^2 + \frac{r^2}{L^2} \mathrm{d}x_1^2 + \frac{r^2}{L^2} \mathrm{d}x_2^2 + \frac{r^2}{L^2} \mathrm{d}x_3^2 + \frac{L^2}{r^2}\mathrm{d}r^2 \nonumber\\
    &+ L^2 \Big[\mathrm{d}\theta^2
+ \cos^2\theta\, \mathrm{d}\psi^2
+ \cos^2\theta\, \sin^2\psi\, \mathrm{d}\phi_1^2
+ \cos^2\theta\, \cos^2\psi\, \mathrm{d}\phi_2^2
+ \sin^2\theta\, \mathrm{d}\phi_3^2
\Big].\label{metricaads5s5}
\end{align}
Below, we study the Polyakov action for the most generic string in the above background.
We  find a particular solution describing the string falling in the radial direction and wrapping the coordinates $\phi_i$.

We write the Polyakov action for the string in AdS$_5\times S^5$, we do not consider fermions, and hence we ignore the Ramond field.
Consider a general embedding,
\begin{equation}\label{generalembedding}
    t=t(\tau,\sigma) , \quad r=r(\tau,\sigma), \quad x_i=x_i(\tau,\sigma), \quad \phi_i= \phi_i(\tau,\sigma), \quad \theta = \theta(\tau,\sigma), \quad \psi= \psi(\tau,\sigma).
\end{equation}
 Denoting $\dot{X}^\mu=\partial_\tau X^\mu$ and $(X^\mu)'=\partial_\sigma X^\mu$, and setting the worldsheet metric $\eta^{\alpha\beta}=\text{diag}(-1,1)$, the Polyakov action reads,
\begin{align}
    &S_{P} = T_{F1} \int \mathrm{d}\sigma \mathrm{d}\tau ~{\cal L}_P= \; T_{F1} \int d\sigma d\tau ~\eta^{\alpha\beta}G_{\mu\nu} \partial_\alpha X^\mu \partial_\beta X^\nu \\
    &\mathcal{L}_P = G_{tt} (t^{\prime \, 2 } - \dot{t}^2) +  G_{rr}(r^{\prime \, 2} - \dot{r}^2) + G_{xx}( x^{\prime \, 2} -  \dot{x}^2)\nonumber \\
    & \qquad \quad + G_{\theta\theta}(\theta^{\prime \, 2} -\dot{\theta}^2 ) + G_{\psi\psi}(\psi^{\prime \, 2}-\dot{\psi}^2 ) + \sum_{i=1}^3 G_{\phi_i \phi_i}(\phi_i^{\prime \, 2} - \dot{\phi}_i^2)\, .\nonumber
\end{align}
The Virasoro constraints are,
\begin{align}
    &T_{\tau\tau} = T_{\sigma \sigma}=\frac{1}{2}\Bigg( G_{tt} (t^{\prime \, 2} + \dot{t}^2) + G_{rr}( r^{\prime \, 2} + \dot{r}^2 ) + G_{xx}(x^{\prime \, 2} + \dot{x}^2 ) \nonumber \\
    & \qquad \qquad \qquad \quad + G_{\theta\theta} (\dot{\theta}^2 + \theta^{\prime \, 2} ) + G_{\psi\psi}( \dot{\psi}^2 +\psi^{\prime \, 2} ) + \sum_{i=1}^3 G_{\phi_i \phi_i}(\phi_i^{\prime \, 2} + \dot{\phi}_i^2) \Bigg) =0 \label{constraint1}\\
    &T_{\tau \sigma }= T_{\sigma \tau } = G_{tt}\dot{t}t^{\prime } + G_{rr}\dot{r}r^\prime  + G_{xx}\dot{x}x^\prime  +  G_{\theta\theta} \dot{\theta}  \theta^{\prime }  + G_{\psi\psi} \dot{\psi} \psi^{\prime }  + \sum_{i=1}^3 G_{\phi_i \phi_i} \dot{\phi}_i \phi_i^{\prime } =0.\label{constraint2}
\end{align}
The equations of motion are 
\begin{align}
    &- \partial_\tau (G_{tt}\dot{t}) + \partial_\sigma ( G_{tt }t^{\prime} ) =0,\label{eqtt}\\
    &- 2 \partial_\tau (G_{rr}\partial_\tau r) + 2 \partial_\sigma (G_{rr} \partial_\sigma r) = \frac{\partial G_{tt}}{\partial r} \eta^{\alpha \beta }\partial_\alpha t \partial_\beta t +~~~ \frac{\partial G_{rr}}{\partial r} \eta^{\alpha \beta }\partial_\alpha \, r \partial_\beta \, r +\nonumber\\
    & ~~\sum_{i=1}^3  \frac{\partial G_{x1x1}}{\partial r} \eta^{\alpha \beta }\partial_\alpha \, x^i \partial_\beta \, x^i       ,\label{eqrr}\\
    &- \partial_\tau (G_{xx}\dot{x}) + \partial_\sigma ( G_{xx }x^{\prime} ) =0, \label{eqxx}\\
    & -2 \partial_\tau (G_{\theta \theta} \partial_\tau \theta) + 2 \partial_\sigma (G_{\theta \theta }\partial_\sigma \theta ) = \frac{\partial G_{\psi \psi }}{\partial\theta} (\psi^{\prime \, 2} - \dot{\psi}^2) + \sum_{i=1}^3 \frac{\partial G_{\phi_i \phi_i}}{\partial \theta }(\phi_i^{\prime \, 2} - \dot{\phi}_i^2),\label{eqthetatheta}\\
    & -2 \partial_\tau (G_{\psi \psi} \partial_\tau \psi ) + 2 \partial_\sigma (G_{\psi \psi } \partial_\sigma \psi ) = \sum_{i=1}^3 \frac{\partial G_{\phi_i \phi_i}}{\partial \psi }(\phi_i'^2 -\dot{\phi}_i^2),\label{eqpsipsi}\\
    &-\partial_\tau (G_{\phi_i \phi_i } \partial_\tau \phi_i ) + \partial_\sigma (G_{\phi_i \phi_i } \partial_\sigma \phi_i) = 0.\label{eqphiphi}
\end{align}
The components of the spacetime metric $G_{\mu\nu}$ are read from eq.(\ref{metricaads5s5}). A consistent truncation of the above partial differential equations and constraints is given by,
\begin{equation}
    t=t(\tau) ,~~ x_j=\text{constant},~~ r=r(\tau), \quad \phi_i(\tau,\sigma) = m_i \sigma, ~~\theta=\theta_0= \text{constant}, ~~ \psi=\psi_0 = \text{constant} \, .\label{tuncationstr}
\end{equation}
The reader can check that on this consistent truncation, eqs.(\ref{constraint2}) and (\ref{eqxx}) are automatically satisfied. Similarly, eq.(\ref{eqpsipsi}) imposes $m_1^2=m_2^2$. Together with this, eq.(\ref{eqthetatheta}) demands (for generic  constant angles $\theta_0,\psi_0$) that $m_3^2=m_1^2=m_2^2=m^2$. On the other hand, on the truncation (\ref{tuncationstr}), eq.(\ref{eqphiphi}) is a wave equation, solved by linear functions $\phi_i=m_i\sigma$.
The interesting equations are (\ref{constraint1}), (\ref{eqtt}) and (\ref{eqrr}). These give the following relations,
\begin{eqnarray}
& & \dot{t}(\tau)=\frac{\hat{E} L^2}{r(\tau)^2},~~~\dot{r}(\tau)^2=\hat{E}^2- m^2 r(\tau)^2 ,~~~\ddot{r}(\tau)-\frac{\dot{r}(\tau)^2}{r(\tau)} +\frac{\hat{E}^2}{r(\tau)}=0.\label{ecuaciones}  
\end{eqnarray}
We defined $\hat{E}$ as the integration constant of the eq.(\ref{eqtt}). The equations (\ref{ecuaciones}), together with the initial condition at $r(\tau=0)=r_{\text{UV}}$ and the choice of time-origin, are solved by
\begin{eqnarray}
& & r(\tau)= r_{\text{UV}}\cos(m \tau),~~~  t(\tau)=\frac{\hat{E}L^2}{m r_{\text{UV}}^2}\tan(m\tau),\nonumber\\
& & \text{or equivalently}~~~r(t)= \frac{r_{\text{UV}}}{\sqrt{1+\left( \frac{m ~r_{\text{UV}}^2 }{\hat{E}~ L^2}\right)^2 t^2}}.\label{solucionultima}
\end{eqnarray}
Let us now calculate the spread Krylov complexity. It is convenient for this, to express things in Nambu-Goto form, we do this next.
\subsection{Krylov spread complexity}
To show the similarity with a particle falling in AdS$_5$ and the treatment of \cite{Caputa:2024sux}, we write the Nambu-Goto action for the string on the embedding in eq.(\ref{tuncationstr}). We need
the induced metric $g_{\alpha\beta}=G_{\mu\nu}\partial_{\alpha}X^\mu\partial_\beta X^\nu$. We find,
\begin{align}
    &g_{\tau\tau} = G_{tt} \dot{t}^2 + G_{rr} \dot{r}^2 \qquad g_{\sigma \sigma } = \sum_{i=1}^3 G_{\phi_i \phi_i} m_i^2 =  L^2 m^2 \qquad g_{\sigma \tau} =0,\nonumber\\
    &S_{NG} = T_{F1}\,  L~ m\,  L_{\sigma} \int \mathrm{d}\tau \; \sqrt{\frac{r^2}{L^2}\dot{t}^2 - \frac{L^2}{r^2}\dot{r}^2}.~~~~\text{we defined}~~L_\sigma=\int d\sigma.\label{NGST}
\end{align}
This is the same action as seen in the case where a particle falls in AdS. 
In the $\tau$-parametrised Nambu-Goto description the canonical Hamiltonian vanishes due to reparametrisation invariance.  The conserved quantity is $P_t$, the target space energy, equivalent to $\hat{E}$ in the Polyakov action.
One can rewrite eq.(\ref{NGST}) as
\begin{equation}
 S_{NG} = T_{F1}\,  m\,~L~  L_{\sigma} \int \mathrm{d}t \; \sqrt{\frac{r^2}{L^2} - \frac{L^2}{r^2}\left(\frac{\dd r}{\dd t}\right)^2}.   
\end{equation}
Which makes clear that this is the same problem as that studied in \cite{Caputa:2024sux}, for a particle of mass\\$M= T_{F1}m~L ~ L_{\sigma}$. The mass of the 'effective particle' is related to the string tension, the AdS radius, the length of the $\sigma$-coordinate and the winding number $m$. The complexity is proportional of the form ${\cal C}(t)\sim t^2$ for all times.
It is interesting to note that in this example we need not appeal to the treatment using the Routhian described in the previous sections. This is because in this case it is possible to have vanishing initial velocities and we do not have a Noether-conserved quantity that needs to be fixed. Our conserved quantity, the winding number does not imply the need of an initial velocity. Different string solutions to eqs.(\ref{constraint1})-(\ref{eqphiphi}) may present Noether conserved quantities and the need to appeal to the Routhian treatment. This is discussed in \cite{ChHNRS}.
\\
Let us close this section by emphasising the main lesson of the string example. Starting from the full Polyakov description in AdS$_5\times S^5$, we identified a consistent truncation describing a string that falls in the holographic direction while wrapping internal angular directions. On this configuration, the Nambu--Goto action reduces to that of an effective massive particle in AdS, with an effective mass fixed by the fundamental-string tension, the AdS radius, the length of the $\sigma$-direction and the winding number. The holographic proper-momentum prescription therefore reproduces the familiar quadratic growth ${\cal C}(t)\sim t^2$, now for a genuine extended string excitation of ${\cal N}=4$ SYM rather than for a pointlike probe. This example also clarifies an important conceptual point: winding is a conserved topological datum, but it does not force non-vanishing initial velocity and therefore does not require the fixed-charge Routhian treatment used in the D0 and D3 examples. More general string embeddings, carrying Noether charges associated with target-space isometries, are expected to combine the two mechanisms discussed above. We now collect these lessons in the conclusions, where we summarise the role of genuine stringy probes, conserved charges, and Routhian dynamics in holographic spread complexity.

\section{Conclusions and closing comments}
In this paper we have given a top-down string-theoretic realisation of holographic Krylov spread complexity.  The proper-momentum proposal of \cite{Caputa:2024sux}, together with the refinements and extensions in \cite{Fan:2024iop,He:2024pox}, was not treated as a phenomenological point-particle rule, but was embedded in genuine objects of type IIA and type IIB string theory.  The three examples studied here---a D0 brane in the ABJM background, a rotating non-BPS D3 brane in $AdS_5\times S^5$, and a wound fundamental string in $AdS_5\times S^5$---show that the falling-particle picture is only the simplest representative of a broader class of holographic probes whose Krylov growth is controlled by radial motion, charges, worldvolume data and internal stringy excitations.

The D0-brane example places the original falling-probe computation on particularly firm ground.  The probe is a bona-fide type IIA excitation and has a natural interpretation in the ABJM theory in terms of dressed monopole operators, following the standard monopole-operator literature \cite{Benna:2009xd,Berenstein:2009sa}.  In the purely radial sector the proper momentum reproduces the expected short-time behaviour, $\dot {\cal C}\sim t$, and therefore ${\cal C}\sim t^2$.  The regulated monopole two-point function provides a complementary boundary derivation; it gives a survival amplitude whose moments determine the Lanczos coefficients and hence the Krylov wavefunctions, in the standard spread-complexity framework \cite{Balasubramanian:2022tpr,Caputa:2023vyr}. {The inclusion of radial fluctuations, which correspond to deforming the classical trajectory upon a fluctuation of the initial radial position, allowed us to calculate small corrections to the rate of change of complexity. These corrections are interpreted as a response of ${\cal C}$ under a change of the initial state. As discussed, the fluctuations studied here are not expressing genuine excitations of dynamical internal degrees of freedom, for which one has to consider the probe D0 fluctuating along the $\mathbb{CP}^3$, which we postpone for future work \cite{ChHNRS}.} 

A central conceptual point of the paper is the treatment of conserved Noether charges.  If one computes the proper momentum naively in a sector where the probe moves along an isometric direction, a non-zero constant contribution to $\dot {\cal C}$ can appear already at $t=0$, see for example \cite{Nastase:2026lhz, Fatemiabhari:2026rob, Fatemiabhari:2026goj}.  This is in tension with the evenness properties of Krylov complexity emphasised, for example, in \cite{Muck:2026top}.  We proposed that the correct bulk prescription in such a sector is a fixed-charge prescription: one Legendre transforms with respect to the cyclic coordinate and works with the corresponding Routhian.  The D0-brane and D3-brane examples show that this procedure restores the required short-time behaviour while retaining the physical dependence on the conserved charge.  Thus the Routhian is not merely a useful rewriting of the equations of motion; it is the bulk implementation of working in a definite charge sector of the boundary theory.

The D3 and F1 examples sharpen this interpretation.  The rotating D3 brane extends the discussion to non-BPS extended operators.  Its angular momentum produces a centrifugal barrier and a precise falling criterion; inside the falling regime the fixed-charge Routhian again gives a rate of complexity compatible with Krylov expectations.  In contrast, the wound fundamental string reduces to an effective massive particle, with an effective mass determined by the string tension, the AdS radius, the length of the $\sigma$-direction and the winding number.  Winding is a conserved topological datum, but it does not force a non-zero initial velocity and therefore does not require the Routhian treatment.  This cleanly separates two mechanisms: Noether charges should be fixed by a Legendre transform, while winding data enter by renormalising the effective mass of the falling excitation. A string with winding and Noether charges implying non-zero initial velocity is studied in \cite{ChHNRS}.

These results connect naturally with previous holographic studies of Krylov complexity in conformal and confining quiver theories \cite{Fatemiabhari:2025poq,Fatemiabhari:2025usn,Fatemiabhari:2026goj} and with the analysis of charged, composite and extended probes in \cite{Nastase:2026lhz,ChHNRS}.  The emerging picture is that the point-particle result captures the collective radial sector, while a full string-theoretic probe carries further information.  A useful way to organise this structure is
\begin{equation}
{\cal C}(t)={\cal C}_{\rm collective}(t)+{\cal C}_{\rm charges}(t)+{\cal C}_{\rm fluctuations}(t)+{\cal C}_{\rm mix}(t).\nonumber
\end{equation}
Here the first term is the proper-momentum contribution associated with radial infall, the second encodes fixed Noether sectors, the third captures worldvolume and internal excitations, and the last represents their possible interactions.

There are several promising directions that this works suggests.  Firstly, the Routhian prescription should be compared directly with symmetry-resolved and charge-resolved Krylov complexity \cite{Caputa:2025mii,Caputa:2025ozd}; one may ask whether different superselection sectors correspond to distinct radial momenta, distinct spectral measures or even distinct Krylov geometries.  Secondly, extended probes with many collective coordinates and fluctuation modes may require a Krylov lattice, or a multi-seed Krylov construction, rather than a single one-dimensional chain, in the spirit of \cite{Das:2024tnw,Craps:2024suj}.  Thirdly, it would be very interesting to extend the present analysis to baryon vertices, giant gravitons, surface and volume defects, and branes in confining or finite-temperature backgrounds.  In such systems the competition between radial motion, worldvolume topology, defect data and charge sectors could provide a refined diagnostic of confinement, screening and thermalisation.  
Finally, one can turn the logic around and pose an inverse problem: to what extent can the Lanczos data of a boundary state reconstruct the bulk effective potential, the conserved charges, or even the stringy nature of the excitation?  The examples studied here suggest that holographic Krylov complexity is not only a measure of growth in Hilbert space, but also a sharp probe of the microscopic string-theoretic origin of the state.

It would also be interesting to relate our paper with the recent works \cite{Baume:2026jyt, Graef:2026pzv}. The recent work \cite{Baume:2026jyt} provides a field-theoretic perspective on the relation between tunable dynamics, spectral chaos and Krylov growth in four-dimensional SCFTs. In particular, their controlled spin-chain subsectors in orbifolds of ${\cal N}=4$ SYM sharpen the question of when Krylov complexity faithfully tracks chaos, a theme related to the holographic probe constructions analysed in the present work. Similarly, \cite{Graef:2026pzv}
treatment of BPS states could be applied to our D3 brane and our string solitons. The exact BPS character of certain states and their Krylov triviality is in correspondence with our smearing and regulation of the state in eq.(\ref{survivalt}) and  the $(\gamma-2)$ factor in eq.(\ref{complexityshorttau}). It would be interesting to make these connections clear.

\section*{Acknowledgements} We want to thank various colleagues for their input that improved the contents and presentation of this work. In particular we thank: Ali Fatemiabhari, Wolfgang M\"uck, Horatiu Nastase, Dibakar Roychowdhury, Joan Sim\'on, Javier Subils. D.C. has been supported by the STFC consolidated grand ST/Y509644-1. M.H. has been supported by the STFC consolidated grant ST/Y509644/1. 
CN is supported by  STFC’s grants ST/Y509644- 1, ST/X000648/1 and ST/T000813/1. The work of R.T. has been supported by EPSRC Grant EP/Z535175/1 and STFC grant UKRI1787.
The work of AVR has received financial support from the Xunta de Galicia (CIGUS
Network of Research Centres and grant ED431C-2021/14), the European Union, the
Mar\'\i a de Maeztu grant CEX2023-001318-M funded by MICIU/AEI/10.13039/501100011033
and the Spanish Research State Agency (grant PID2023-152148NB-I00).

\bibliographystyle{JHEP}
\bibliography{main.bib}

\end{document}